\documentclass{elsart}
\usepackage{lineno}

\usepackage{natbib}
\usepackage{graphicx}  

\usepackage{amssymb}

\begin{document}

\begin{frontmatter}



\title{The Offline Software Framework of the Pierre Auger Observatory}


\author[torino]{S. Argir\`o},
\author[cbpf]{S.L.C. Barroso\thanksref{sergio}},
\author[neu]{J. Gonzalez\thanksref{nsf}},
\author[unam]{L. Nellen\thanksref{lukas}},
\author[neu]{T. Paul\thanksref{nsf}\corauthref{tom}},
\author[lsu]{T.A. Porter\thanksref{doe}},
\author[unicamp]{L. Prado Jr.\thanksref{sergio}},
\author[fzk]{M. Roth\thanksref{markus}},
\author[fzk]{R. Ulrich}, and
\author[ngp]{D. Veberi\v{c}\thanksref{darko}}

\corauth[tom]{corresponding author: \texttt{t.paul@neu.edu}}

\thanks[sergio]{The work of S.L.C.B. and L.P.Jr was supported by the Brazilian funding agencies CNPq and FAPESP.}
\thanks[nsf]{The work of T.P. and J.G. was supported in part by the US National Science Foundation.}          
\thanks[lukas]{The work of L.N. was supported by CONACyT, DGAPA-UNAM (PAPIIT programme), and CIC-UNAM.}
\thanks[doe]{The work of T.A.P. was supported in part by the US Department of Energy.}
\thanks[markus]{The work of M.R. was supported by the HHNG-128 grant of the German Helmholtz association.}
\thanks[darko]{The work of D.V. was supported by the Slovenian research agency ARRS.}

\address[torino]{INFN and University of Torino, Via P. Giuria 1, I-10125 Torino, Italy}
\address[cbpf]{Centro Brasileiro de Pesquisas F{\'\i}sicas, Rua Dr. Xavier Sigaud, 150, Rio de Janeiro-RJ, CEP 22290-180, Brazil}
\address[neu]{Northeastern University, 360 Huntington Ave., Boston, MA, 02115 USA}
\address[unam]{Departamento de F{\'\i}sica de Altas Energ{\'\i}as, Instituto de Ciencias Nucleares, Universidad Nacional Autonoma de M{\'e}xico, M{\'e}xico D.F., C.P 04510}
\address[lsu]{Louisiana State University, Baton Rouge, LA, 70803, USA and Santa Cruz Institute for Particle Physics, University of California, Santa Cruz, CA, 95046, USA}
\address[unicamp]{Instituto de F{\'\i}sica Gleb Wataghin, Universidade Estadual de Campinas UNICAMP, Campinas-SP, CP 6165 CEP 13083-970, Brazil}
\address[fzk]{Karlsruhe Institute of Technology KIT, University and Forschungszentrum Karlsruhe, POB 3640, D-76021 Karlsruhe, Germany}
\address[ngp]{University of Nova Gorica, Vipavska 13, PO Box 301, SI-5001, Nova Gorica, Slovenia}

\begin{abstract}
The Pierre Auger Observatory is designed to unveil the nature and the
origins of the highest energy cosmic rays. The large and geographically
dispersed collaboration of physicists and the wide-ranging collection of simulation
and reconstruction tasks pose some special challenges for the offline
analysis software.
We have designed and implemented a general purpose framework which allows 
collaborators to contribute algorithms and sequencing instructions to build up the
variety of applications they require.  The framework includes
machinery to manage these user codes, to organize the abundance of
user-contributed configuration files, to facilitate multi-format file
handling, and to provide access to event and time-dependent detector
information which can reside in various data sources.  A number of
utilities are also provided, including a novel geometry package which
allows manipulation of abstract geometrical objects independent of
coordinate system choice. The framework is implemented in C++, 
and takes advantage of object oriented design and common open source
tools, while keeping the user side simple enough for C++ novices to
learn in a reasonable time.
The distribution system incorporates unit and acceptance
testing in order to support rapid development of both the core
framework and contributed user code.  
\end{abstract}


\end{frontmatter}

\section{Introduction}
The offline software framework of the 
Pierre Auger Observatory~\cite{Abraham:2004dt} 
provides an infrastructure to support a variety of distinct 
computational tasks
necessary to analyze data gathered by the observatory.
The observatory itself is designed to measure the
extensive air showers produced by the highest
energy cosmic rays ($> 10^{19}$~eV) with the goal 
of discovering their origins and composition.  
Two different techniques are used to detect
air showers.  Firstly, a collection of telescopes
is used to sense the fluorescence light produced
by excited atmospheric nitrogen as the cascade of 
particles develops and deposits energy in the 
atmosphere.
This method can be used only when the sky is moonless and dark, 
and thus has roughly a 15\% duty cycle.
Secondly, an array of 
detectors on the ground is used to sample particle 
densities and arrival times as the air shower impinges upon the Earth's surface.
Each surface detector consists of a tank containing 12 tons of
purified water instrumented with photomultiplier tubes to detect
the Cherenkov light produced by passing particles.  
The surface detector has a 100\% duty cycle.
A subsample of air showers detected by both instruments,
dubbed hybrid events, are very precisely measured and
provide an invaluable energy calibration tool.
In order to provide full sky coverage, the baseline design of the observatory calls for 
two sites, one in the southern hemisphere and one in the north.
The southern site is located in Mendoza, Argentina, and construction
there is nearing completion, at which time the observatory 
will comprise 24 fluorescence telescopes overlooking 
1600 surface detectors spaced 1.5~km apart on a
hexagonal grid.  The state of Colorado in the USA 
has been selected as the location for the northern site.

The requirements of this project place rather strong demands
on the software framework underlying data analysis.  Most importantly,
the framework must be flexible and robust 
enough to support the collaborative effort of a large
number of physicists developing a variety of applications over
the projected 20 year lifetime of the experiment. 
Specifically, the software must 
support simulation and reconstruction of events using
surface, fluorescence and hybrid methods, as well as
simulation of calibration techniques
and other ancillary tasks such as data preprocessing.  
Further, as the experimental
run will be long, it is essential that the software be
extensible to accommodate future upgrades to the observatory
instrumentation.  The offline framework 
must also handle a number of data formats in order to deal with
event and monitoring information from a variety of instruments, 
as well as the output of air shower simulation codes.
Additionally, it is desirable that all physics code
be ``exposed'' in the sense that any collaboration member must be able to 
replace existing algorithms with his or her own in a straightforward
manner.  Finally, while the underlying framework itself may
exploit the full power of C++ and object-oriented design, 
the portions of the code directly used by physicists 
should not assume a particularly detailed knowledge of
these topics.  

The offline framework was designed with these principles 
in mind.  Implementation began in 2002, and the first 
physics results based upon this code were presented at 
the $29^{\rm th}$ International Cosmic Ray Conference~\cite{icrc2005}.
  

\section{Overview}

\begin{figure}[t!]
\centering
\includegraphics[width=3.6in]{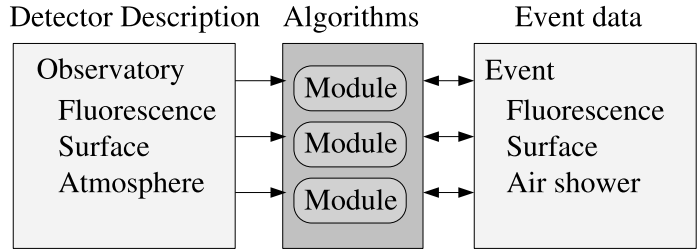}
\caption{General structure of the offline framework.
Simulation and reconstruction tasks are broken down
into modules.  Each module is able to read information
from the detector description and/or the event,
process the information, and write the results back
into the event.}
\label{f:general}
\end{figure}

The offline framework comprises three principal parts:
a collection of processing {\em modules} which can
be assembled and sequenced through instructions 
provided in an XML file, an {\em event} data model
through which modules can relay data to one another
and which accumulates all simulation and reconstruction
information, and a {\em detector description} which provides
a gateway to data describing the configuration and 
performance of the observatory as well as atmospheric 
conditions as a function of time.  These
ingredients are depicted in Fig.~\ref{f:general}.

This approach of pipelining processing modules which communicate through
an event serves to separate data from the algorithms which operate on 
these data.  Though this approach is not particularly characteristic 
of object oriented design, it was used nonetheless, as it better satisfies
the requirements of physicists whose primary objective
is collaborative development and refinement of algorithms.

These components are complemented by
a set of foundation classes and utilities for
error logging, physics and mathematical manipulation,
as well as a unique package supporting
abstract manipulation of geometrical objects. 

Each of these aspects of the framework is described in more detail
below.

\section{User Modules and Run Control} \label{sec:config}

Experience has shown that most tasks of interest of the Pierre Auger
Collaboration can be factorized into sequences of self contained
processing steps.  Physicists prepare such processing algorithms
in so-called {\em modules}, which they register with the framework
via a macro.  This modular design allows
collaborators to exchange code easily, compare algorithms and
build up a wide variety of applications by combining modules in various 
sequences. 

Modules inherit a common interface which declares the methods
that carry out processing.  Specifically, module authors
must implement a \texttt{Run} method which is called once per event, as
well as \texttt{Init} and \texttt{Finish} methods to be called at the
beginning and end of a processing job.  Authors invoke a 
macro in the module class declaration which registers a 
factory function used by the framework to instantiate 
the module when requested. The registry mechanism provides a fall-back hook 
that handles requests for modules not known to the registry. 
Dynamical loading using this fall-back mechanism
is currently under development.   
Modules themselves are not instrumented with a means
to place requirements on versions of other modules or 
on module execution order; instead, the configuration machinery described in 
section~\ref{sec:configuration} is used to set such requirements.

For most applications, run-time control over module sequences 
is afforded through a {\em run controller} which invokes the various
processing steps within the modules according to a set of externally
provided instructions.  We have constructed
a simple XML-based~\cite{xml} language for 
specifying these sequencing instructions.  The decision to
employ XML for this task was based on several considerations.
Firstly, sequencing via interpreted XML files allows one
to set up or modify the behavior of a run without compiling,
and so offers the convenience of an interpreted language.
In addition, XML syntax is simple and relatively familiar to the community,
and hence imposes a minimal learning curve compared to
most scripting languages.  
Furthermore, XML sequencing instructions can be  
straightforwardly logged using the same mechanisms employed for 
other types of configuration logging, as described in some detail 
in section~\ref{sec:configuration}.
Finally, XML proves sufficient to support
our commonest applications, which do not require especially 
intricate sequencing control. In fact the majority of 
applications require only two fundamental instructions.
First, a \texttt{module} element
instructs the run controller to invoke the processing method
of the module whose name appears between the begin and end tags.
Second, a \texttt{loop} tag is used to command looping over modules, 
including arbitrarily deep loop nests.  This tag may be
decorated with attributes indicating the number of times 
to loop, and whether or not the contents of the event should 
be pushed onto a stack before the modules contained in the loop
begin acting upon it.  
It is appropriate to push the event to a stack if, for example,
one wishes to repeat a random process starting from the same input.
On the other hand, one would disable the push to stack in order
to implement an iterative procedure distributed over several modules. 
Fig.~\ref{f:xml} shows a simple example 
of the structure of a sequencing file.

\begin{figure}[tbh]
\centering
\begin{small} \begin{verbatim}
   <sequenceFile>
      <loop numTimes="unbounded">
        <module> SimulatedShowerReader </module>
        <loop numTimes="10" pushToStack="yes">
           <module> EventGenerator     </module>
           <module> TankSimulator      </module>
           <module> TriggerSimulator   </module>
           <module> EventExporter      </module>
        </loop>
      </loop>
   </sequenceFile>
\end{verbatim} \end{small}
\caption{Simplified example in which an XML file
sets a sequence of modules to conduct a simulation
of the surface array.  {\tt <loop>} and {\tt <module>}
tags are interpreted by the run controller, which
invokes the modules in the proper sequence.  In this
example, simulated showers are read from a file, and
each shower is thrown onto the array in 10 
random position by an {\tt EventGenerator}.  Subsequent 
modules simulate the response of the surface detectors
and trigger, and export the simulated data to file.  
The \texttt{pushToStack="yes"} attribute instructs the Run
Controller to store the event when entering the 
loop, and restore it to that state when returning back
to the beginning of the loop. 
Note that XML naturally accommodates common sequencing
requirements such as nested loops.}
\label{f:xml}
\end{figure}
Modules can signal the run controller via return codes
of the \texttt{Init}, \texttt{Run} and \texttt{Finish} methods,
and command it to
break a loop or to skip all subsequent modules up
to the next loop tag.


\section{Configuration} \label{sec:configuration}

Parameters, cuts and configuration instructions used by
modules or by the framework itself are stored in 
XML files.  A globally accessible {\em central configurator} points
modules and framework components to the location of 
their configuration data, and creates parsers to assist in reading information 
from these locations.   The locations of configuration data are specified 
in a so-called {\em bootstrap} file, and may comprise local filenames, URIs~\cite{URI} or XPath~\cite{xpath} addresses.
The name of the bootstrap file is passed to the application at run time via the command line.

The configuration mechanism can also concatenate all configuration
data accessed during a run and write it in a log file.
This log file includes a preamble with a format identical 
to that of a bootstrap file, with XPath 
addresses specifying the locations of all the configuration 
data in the file. In this way, a log file can subsequently be read,
as though it were a bootstrap file, in order
to reproduce a run with an identical configuration. 

This configuration logging mechanism may also be used to record the versions
of modules and external libraries which are used during a run.
External package versions are determined at build time by our GNU autotools-based
build machinery~\cite{autotools}, which searches the local system for 
required packages and verifies that versions of
different packages are compatible with one another.  The
build system generates a script which can later be 
interrogated by the logging mechanism to record package versions employed for 
a particular run.  Module versions are declared in the code by module authors,
and are accessible to the configuration logging mechanism through the module interface.

To check configuration files for errors, the W3C XML Schema~\cite{schema} standard
validation is employed throughout. This approach is used 
not only to validate internal framework configuration, 
but also to check configuration files of modules prepared 
by framework users.  This has proved to be successful in saving coding time
for developers and users alike, and facilitates much more detailed
error checking than most users are likely to implement on their own.
Further details on how configuration files are parsed and validated
are given in section~\ref{sec:parsing}.

The configuration machinery is also able to verify configuration file
contents against a set of default files by employing MD5 digests~\cite{md5}. 
The default configuration files are prepared by the framework developers and the analysis 
teams, and reference digests are computed from these files at build time.
At run time, the digest for each configuration file is recomputed 
and compared to the reference value.  Depending on run-time options,
discrepant digests can either force program termination, or can simply
log a warning.
This machinery provides a means for those managing data 
analyses to quickly verify that configurations in use are
the ones which have been recommended for the task at hand.

\section{Data Access}

The offline framework provides two parallel hierarchies for accessing
data: the {\em event} for reading and writing information that changes per event, 
and the read-only {\em detector description} for retrieving static or 
relatively slowly varying information such as detector geometry, calibration constants,
and atmospheric conditions.  

\subsection{Event} \label{sec:event}

\begin{figure}[htb]
\centering
\includegraphics[width=3.6in]{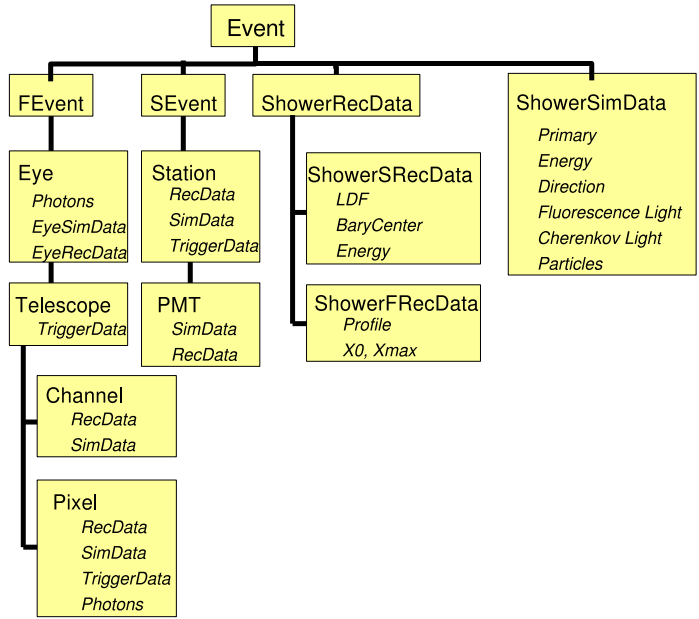}
\caption{Hierarchy of the event interface.
The top level Event encapsulates objects representing 
Fluorescence and Surface events (\texttt{FEvent} and
\texttt{SEvent} respectively), as well as reconstructed
and simulated shower data (\texttt{ShowerRecData} and
\texttt{ShowerSimData} respectively).  These components
are further subdivided into objects representing simulated,
reconstructed and triggering data at the level of 
individual telescopes, tanks and photomultiplier tubes.
}
\label{f:event}
\end{figure}

The {\em Event} data model contains all raw, calibrated, 
reconstructed and Monte Carlo data and acts as the 
principal backbone for communication between modules.
The overall structure comprises a collection
of classes organized following the hierarchy normally 
associated with the observatory instruments, with
further subdivisions for accessing such information as  Monte Carlo truth, 
reconstructed quantities, calibration information and raw data.
A non-exhaustive illustration of this hierarchy is given in 
Fig.~\ref{f:event}.  
User modules access the event through
a reference to the top of the hierarchy which is 
passed to the module interface by the run controller.

Since the event constitutes the inter-module communication backbone, 
reference semantics are used
throughout to access data structures in the event, and constructors are 
kept private to prevent accidental copying of event components.  
For example, to retrieve the object representing
the first photomultiplier tube (PMT) object of station number 157
belonging to the surface detector (S) portion of the event, 
one could write simply,
\begin{small} \begin{verbatim}
PMT& pmt1 = theEvent.GetSEvent().GetStation(157).GetPMT(1);
\end{verbatim} \end{small}
where \texttt{theEvent} is a reference to the top of the event
hierarchy.

The event is built up dynamically as needed, and 
is instrumented with a simple protocol allowing modules to interrogate the
event at any point to discover its current constituents.
This protocol provides the means for a given module to 
determine whether the input data required to carry
out the desired processing is available.   
As an example, consider the case of the Monte Carlo truth
belonging to a PMT object called \texttt{thePMT}.
Attempting to access a non-existent subcomponent
\texttt{thePMT} raises an exception:
\begin{small} \begin{verbatim}
PMT& iDontExistYet = thePMT.GetSimData();// exception
\end{verbatim} \end{small}

Checking for the existence of the desired data,
creating an event subcomponent, and retrieveing a 
handle to the data therein would be carried out 
as follows:
\begin{small} \begin{verbatim}
if (!thePMT.HasSimData())                // check for SimData
  thePMT.MakeSimData();                  // create SimData
PMT& thePMTSim = thePMT.GetSimData();    // success
\end{verbatim} \end{small}

The structure of the event interface cannot be modified by modules.
While this restriction limits the flexibility available to module 
authors, it does facilitate straightforward module interchangeability, which,
as discussed in section~\ref{sec:examples}, is 
of primary importance in our case.  In practice, when 
users find the event structure does not accommodate their 
needs, they may implement ad hoc inter-module communication as temporary
solution, and propose the required event interface changes to the 
framework developers.
Though this approach does require periodic developer intervention, it 
has not proved to be overly problematic for our project.  

It is worth noting that the use of an event data model as an algorithm 
communication backbone is not an uncommon approach, and is employed by 
other high energy physics and astrophysics 
experiments~\cite{d0framework,cms,icetray}.
In our case, however, the data access methods are somewhat
less generalized than the techniques employed by larger
experiments (see for example \cite{cms}).  Our feeling is that
the cost incurred in terms of flexibility is reasonably offset by 
user-side simplicity.

\begin{figure}[htb]
\centering
\includegraphics[width=3.6in]{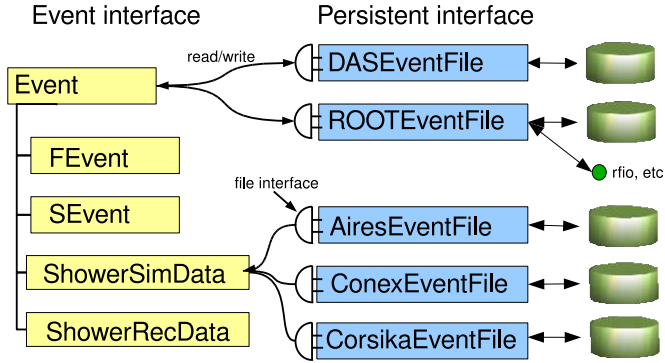}
\caption{Event input/output.  The section labeled ``Event Interface''
portrays a subset of the hierarchy depicted in Fig.~\ref{f:event}.
Data are transferred between this transient event and 
persistent objects through a common file interface.  
Different file
implementations are able to read and/or write
in different formats, including those used by the 
data acquisition systems (\texttt{DAS} formats), 
formats used by other simulation
packages, as well as a ``native'' format (\texttt{ROOTEventFile})
which accommodates  all raw data, reconstructed quantities, and Monte Carlo truth.  
}
\label{f:eventBackend}
\end{figure}

The event representation in memory, or transient event, 
is decoupled from the representation on file, or persistent event.  
When a request is made to write
event contents to file, the data are transferred from the 
transient event through
a so-called file interface to the persistent event, 
which is instrumented with machinery for serialization. 
Conversely, when data are requested from file, a file interface
transfers the data from the persistent event to the
appropriate part of the event interface.
Users can transfer all or part of the event from memory
to a file at any stage in the processing, and reload the event and 
continue processing from that point onward.
Various file formats are handled using the
file interface mechanism, including raw event and
monitoring formats as well as the different formats employed by the 
AIRES~\cite{aires}, CORSIKA~\cite{corsika}, CONEX~\cite{conex}
and SENECA~\cite{seneca} air shower simulation packages.
Fig.~\ref{f:eventBackend} contains a diagram of this event
input/output mechanism.

Event serialization is implemented
using the ROOT~\cite{root} toolkit, though 
the decoupling of the transient and persistent events
is intended to allow for relatively straightforward changes of 
serialization machinery if this should ever prove 
to be advantageous. 
When it becomes necessary to modify
the event structure, backward compatibility is provided 
via the ROOT schema evolution mechanism.  
We note that similar strategies for event persistency 
and schema evolution
have been adopted by other large experiments~\cite{atlas-edm}.

\subsection{Detector Description} \label{sec:detector}

\begin{figure}[htb]
\centering
\includegraphics[width=4.2in]{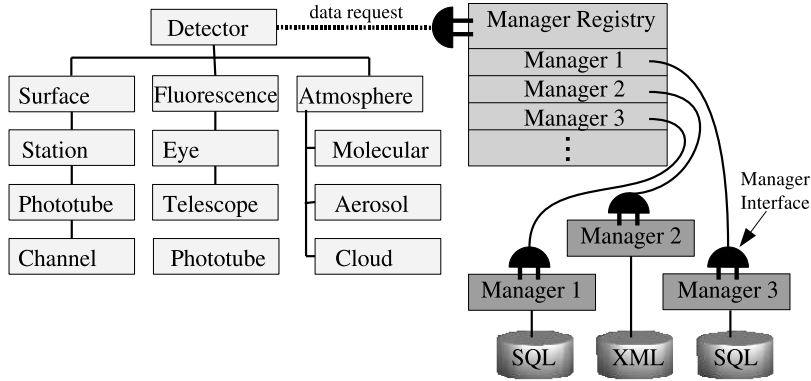}
\caption{
Machinery of the detector description.  The user interface
(left) comprises a hierarchy of objects describing the 
various components of the observatory.  These objects relay
requests for data to registry of managers
(right) which handle multiple data sources and formats.
}
\label{f:detector}
\end{figure}

The {\em detector description} provides a unified interface
from which module authors can retrieve
non-event data including the detector configuration and performance at 
a particular time as well as atmospheric conditions. 
Like the event interface, the 
detector interface is organized following the hierarchy normally 
associated with the observatory instruments, and provides a 
set of simple-to-use access functions to extract data.
Data requests are passed by this interface to a registry
of so-called {\em managers}, each of which is capable of 
extracting a particular sort of information from a 
particular data source.  Data retrieved from a manager
are cached in the interface for future use.  
In this approach, the user deals with a
single interface even though the data sought may reside in 
any number of different sources.  Generally we choose to store
static detector information in XML files, and 
time-varying monitoring and calibration data in MySQL~\cite{mysql}
databases.  However, as the project evolves it sometimes 
happens that access to detector data in some other format is required,
perhaps as a stop-gap measure.  The manager mechanism allows
one to quickly provide simple interfaces in such cases, keeping
the complexity of accessing multiple formats hidden from the user.
The structure of the detector description 
machinery is illustrated in Fig.~\ref{f:detector}.

Note that it is possible to implement more than one manager
for a particular sort of data.  In this way, a special 
manager can override data from a general manager.  For 
example, a user can decide to use a database for the
majority of the description of the detector, but override some data
by writing them in an XML file which is interpreted by 
the special manager.  The specification of which data 
sources are accessed by the manager registry and in 
what order they are queried is detailed in a configuration
file.  The configuration of the manager registry is transparent
to the user code.

The detector description is also equipped to support a 
set of plug-in functions, called {\em models} which can be used for 
additional processing of data retrieved through the detector interface.
These are used primarily to interpret atmospheric
monitoring data.  As an example, users can invoke
a model designed to evaluate attenuation of light due to aerosols 
between two points in the atmosphere.  This model
interrogates the detector interface to find the 
atmospheric conditions at the specified time, and
computes the attenuation.  Models can also be 
placed under command of a {\em super-model} which
can attempt various methods of computing the desired
result, depending on what data are available for the 
specified time.

The manager mechanism has also proved convenient
for generating detector data for use by specialized 
Monte Carlo simulations.  For instance, when 
studying reconstruction techniques, it is sometimes
useful to include hypothetical surface detector stations 
in a simulation run, where the positions of the hypothetical
stations are defined in the reference frame of the shower being simulated.  
A manager has been prepared which uses the simulated 
shower geometry to project a user-defined collection of 
hypothetical stations onto the ground,  overlaying 
them on the actual station positions.  Both sorts of 
station information are accessible transparently 
via the same interface.

\section{Utilities}

The offline framework is built on a collection of 
utilities, including an XML parser, an 
error logger, various mathematics and physics services
including a novel geometry package, 
testing utilities and a set of foundation classes to represent
objects such as signal traces, tabulated functions and
particles.  
In this section, we describe the parsing and geometry packages
in more detail.

\subsection{XML Parsing and Validation} \label{sec:parsing}

As noted previously, XML is employed to store configuration
data for framework components and user contributed modules.
Our XML reading utility, named simply {\em reader},
is built upon the Xerces~\cite{xerces} validating parser,
and is designed to provide ease-of-use at the expense of somewhat reduced 
flexibility compared to the Xerces APIs.   
The reader utility also provides additional features such as units handling for dimensional quantities, 
expression evaluation, and casting of data in XML files to atomic types, 
Standard Template Library~\cite{stl} containers, and data types
specific to the Auger Observatory software.
Navigation through hierarchical data is supported by a handful of methods for 
locating children and siblings of a given XML element.  Data casting is provided 
through overloaded access functions.

For instance, a configuration file can contain a dimensional 
quantity with units specified by an expression in the tag attribute:
\begin{small} \begin{verbatim}
<g unit="meter/second^2"> 9.8 </g>
\end{verbatim} \end{small}
Upon request, the reader casts the data between the \texttt{<g>} tags to the 
data type used in the access function,
evaluates the \verb+meter/second^2+ expression in the attribute, and then 
uses the result to convert the \texttt{9.8} into pre-defined internal units.  

Validation rules are specified using the XML Schema standard~\cite{schema}, which 
is well supported by Xerces and which has proved to be more 
palatable to our user base than the older DTD~\cite{dtd} standard.
The built-in Schema types have been extended with a
collection of data types commonly used by the collaboration, including lists,
three-vectors, and tabulated functions as well quantities which require
an associated unit.  For example, the line of XML shown above 
can be validated using the Schema fragment,
\begin{small} \begin{verbatim}
<xs:element name="g" type="auger:doubleWithUnit"/>
\end{verbatim} \end{small}
where the prefix \texttt{xs} denotes the standard Schema namespace, and 
the \texttt{auger} prefix indicates the namespace containing our extensions
of the standard types.  Here the \texttt{doubleWithUnit} type
specification guarantees that exactly one double precision number 
appears in the element, and that a \texttt{unit} attribute is present in the 
corresponding tag.  
  
One can build up rather involved types in a straightforward
manner. For instance, 
it is useful to have the ability to define functions 
using either tabulated values:

\begin{small} \begin{verbatim}
<EnergyDistribution> 
  <x unit="MeV"> 100 330 1000 </x>
  <y> 1 0.95 0.5 </y>
</EnergyDistribution>
\end{verbatim} \end{small}

or using a parametrization with limits:

\begin{small} \begin{verbatim}
<EnergyDistribution> 
  <PDF> 1/x </PDF>
  <min unit="MeV"> 100 </min>
  <max unit="GeV"> 1 </max>
</EnergyDistribution>
\end{verbatim} \end{small}

The corresponding Schema rules 
require the \texttt{<EnergyDistribution>} 
to be specified by a group of XML elements:

\begin{small} \begin{verbatim}
<xs:element name="EnergyDistribution">
  <xs:complexType>
    <xs:group ref="distributionFunction"/>
  </xs:complexType>
</xs:element>
\end{verbatim} \end{small}

where this group in turn is defined as a choice between two 
sequences of XML elements dictating
the two ways to specify the distribution function mentioned above:

\begin{small} \begin{verbatim}
<xs:group name="distributionFunction">
  <xs:choice>
    <xs:sequence>
      <xs:element name="x" type="auger:listOfDoublesWithUnits"/>
      <xs:element name="y"  type="auger:listOfDoubles"/>
    </xs:sequence>
    <xs:sequence>
      <xs:element name="PDF" type="xs:string"/>
      <xs:element name="min" type="auger:doubleWithUnit"/>
      <xs:element name="max" type="auger:doubleWithUnit"/>
    </xs:sequence>
  </xs:choice>
</xs:group>
\end{verbatim} \end{small}

The combination of XML and XML Schema validation 
enables us to support quite detailed configurations
and robust checking, both in the internal framework
configuration and in user-provided physics modules.   
To give a specific example, we consider the case 
of a so-called ``particle injector''
module which is used to randomly draw particles from 
various distributions and pass them
to downstream detector response simulation modules.  
In configuring this module we wish to support a number of options
in different combinations, including: placing particles 
at specific locations relative to a detector, or distributing them 
over cylindrical or spherical surfaces around the detector; 
selecting different particle types with differing probabilities; 
setting discrete energies or drawing energies from a 
distribution; setting discrete zenith and azimuthal angles
or drawing them from distributions; and 
describing distributions either analytically through 
an expression, or using a tabulated function.
The hierarchical model employed by XML allows one to 
notate all these options in a human-readable form.  Further, XML Schema
validation facilitates detailed policing of the corresponding configuration 
file, so that for instance one can require a distribution to be 
described {\em either} analytically {\em or} in tabular form,
as outlined in the example above.

Note that the general effectiveness of XML and XML Schema in 
software for large experiments
has been noted by other authors~\cite{ice3-xml}.

\subsection{Geometry}

As discussed previously, the Pierre Auger Observatory comprises many
instruments spread over a large area and, in the case of the
fluorescence telescopes, oriented in different directions.
Consequently there is no naturally preferred coordinate system for the
observatory; indeed each detector component has its own natural 
system, as do components of the event such as the air shower itself.
Furthermore, since the detector spans more than 50~km from side to side,
the curvature of the earth cannot generally be neglected.  In such a
circumstance, keeping track of all the required
transformations when performing geometrical computations is tedious
and error prone.

This problem is alleviated in the offline geometry package by
providing abstract geometrical objects such as points and vectors.
Operations on these objects can then be written in an abstract way,
independent of any particular coordinate system. Internally, the
objects store components and track the coordinate system used. There
is no need for pre-defined coordinate system conventions, or
coordinate system conversions at module boundaries. The transformation
of the internal representation occurs automatically.

Despite the lack of a  single natural coordinate system
for the observatory, there are several important
coordinate systems available. A registry mechanism provides access to
a selection of global coordinate systems. Coordinate
systems related to a particular component of the detector, like a
telescope, or systems which depend on the event being processed, such as a shower
coordinate system, are available through access functions belonging to the
relevant classes of the detector or event structures.

Coordinate systems are defined relative to other coordinate
systems. Ultimately, a single root coordinate system is required. It must
be fixed by convention, and in our case we choose an origin at the center 
of the Earth. Other base coordinate systems and a caching mechanism help to avoid the
construction of potentially long chains of transformations when going
from one coordinate system to another. 

The following is a simple example of how the geometry and units
packages are used together:
\begin{small} \begin{verbatim}
Point pos(x*km, y*km, z*km, posCoordSys);
Vector dist(deltaX, deltaY, deltaZ, otherCoordSys);

Point newPos = pos + dist;
cout << "X = " << newPos.GetX(outCoordSys)/m << " meters";
\end{verbatim} \end{small}
The variables \texttt{x}, \texttt{y}, and \texttt{z} are provided by
some external source, in the units indicated (km), whereas \texttt{deltaX},
\texttt{deltaY}, and \texttt{deltaZ} are results from a previous
calculation, already in the internal units. Coordinate systems are
required whenever components are used explicitly. Units are used on
input and output of data and when exchanging information with external
packages.

\begin{figure}[htb]
  \centering
  \includegraphics{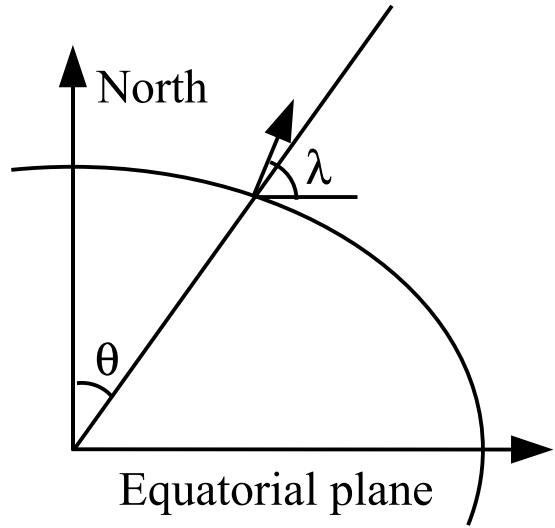}
  \caption{The geodetic latitude \(\lambda\) is defined as the angle
    between the local vertical and a plane parallel to the equatorial
    plane. For an elliptical shape, it is not just the complement of
    the zenith angle \(\theta\) in the definition of spherical
    coordinates.}
  \label{fig:geodesy}
\end{figure}

The surveying of the detector utilizes Universal Transverse Mercator
(UTM) coordinates with the WGS84 ellipsoid. These coordinates are
convenient for navigation. They involve, however, a non-linear,
conformal transformation. The geometry package provides a
\texttt{UTMPoint} class for dealing with positions given in UTM, in
particular for the conversion to and from Cartesian
coordinates. This class also handles the geodetic conventions, which define the
latitude based on the local vertical (see Fig.~\ref{fig:geodesy}), 
as opposed to the angle \(90^\circ -
\theta\), where \(\theta\) is the usual zenith angle in spherical
coordinates. 

The high degree of abstraction makes use of the geometry package 
quite easy. Uncontrolled, repeated
coordinate transformations, though, can be a problem both for
execution speed and for numerical precision. To control
this behavior, it is possible to force the internal representation of
an object to use a particular coordinate system. The geometry package
guarantees that no transformations take place in operation where all
objects are represented in the same coordinate system. This provides a
handle for experts to control when transformations take place.


\section{Build System and Quality Control} \label{sec:build}

To help ensure code maintainability and stability in 
the face of a large number of contributors and 
a decades-long experimental run, unit and acceptance testing
are integrated into the offline framework build and
distribution system. This sort of quality assurance
mechanism is crucial for any software which must continue
to develop over a timescale of many years.

Our build system is based on the GNU autotools~\cite{autotools},
which provide hooks for integrating tests with the 
build and distribution system.  A substantial collection
of unit tests has been developed, each of which
is designed to comprehensively exercise a single framework
component.  We have employed the CppUnit~\cite{cppunit}
testing framework as an aid in implementing these unit tests.
In addition to such low-level tests, a collection of higher-level
acceptance tests has been developed which is used to verify 
that complete applications continue to function properly 
during ongoing development.
Such acceptance tests typically run full physics applications
before and after each code change and notify developers in case 
of any unexpected differences in results.

As a distributed cross-platform project, the Auger Offline
software must be regularly compiled and checked on numerous platforms.
To automate this process, we have employed the tools
provided by the BuildBot project~\cite{buildbot}.  The 
BuildBot is a Python-based system in which a master daemon
is informed each time the code repository has been significantly
altered.  The master then triggers a collection of build slaves running
on various platforms to download the latest code, build it, run 
the unit and acceptance tests, and inform the appropriate developers
in case problems are detected.  This has proved to be a very effective
system, and provides rapid feedback to developers in case of problems.

\section{External packages}

The choice of external packages upon which to build the offline
framework was dictated not only by package features
and the requirement of being open-source, but also by our
best assessment of prospects for longevity.  At the same time,
we attempted to avoid locking the design to any single-provider
solution.  To help achieve this, the functionality of 
external libraries is often provided 
to the client code
through wrappers or fa{\c c}ades, as in the case of XML 
parsing described in sections~\ref{sec:config} and~\ref{sec:parsing}, or through 
a bridge, as in the case of the detector description described
in section~\ref{sec:detector}.  The collection of external
libraries currently employed includes ROOT~\cite{root} for
serialization, Xerces~\cite{xerces} for XML parsing and validation,
CLHEP~\cite{clhep} for expression evaluation and geometry foundations,
Boost~\cite{boost} for its many valuable C++ extensions, and optionally
Geant4~\cite{g4} for detailed detector simulations.


\section{Examples}\label{sec:examples}

In this section we demonstrate the application of the offline software to typical
simulation and reconstruction user tasks.
We consider specifically the case of {\em hybrid} simulation and reconstruction, which
involves combining the sequences for surface and fluorescence detector simulation 
and reconstruction. The discussion is meant to illustrate some of the advantages
of modularization at the level of algorithms, as well as the simple XML-based
sequencing control.

\subsection{Hybrid Detector Simulation}

Simulation of events observed by the hybrid detector typically involves
creation of a shower using a Monte Carlo generator~\cite{aires,corsika, conex, seneca}, 
the simulation of the response and
triggering of the surface array to the particles arriving at the ground, and the
simulation of the telescope response and triggering to the profile
of fluorescence light emitted along the shower track.
Finally, event building and export to various data formats
can be performed.

Detector simulations can be broken down into a sequence of steps, each
of which is generally encapsulated within a separate module.
For example, simulation of the surface detector typically begins with 
a module which places the simulated shower impact point somewhere within the 
surface array configuration.  This is followed by a module which uses this information
to determine which particles enter into which water tanks.  Subsequent modules
then simulate the particle energy loss and Cherenkov light
emission in each tank, the response of the phototubes and tank electronics, and the
local tank trigger.  A final module simulates the response of the central 
trigger, which considers information
from multiple detector components when determining whether to record the event.
Simulation of a fluorescence event involves modules for placing the simulated
shower at some distance from one of the eyes, simulating the fluorescence and 
Cherenkov light emitted by the shower as it develops, and finally
simulating the response of the fluorescence telescopes, electronics and
triggering.
In Fig.~\ref{f:HSim} we show a typical module sequence for hybrid detector
simulation.  Each {\tt <module>} element designates an individual simulation step.

\begin{figure}[htb]
\footnotesize
\centering
\begin{small} \begin{verbatim}
<sequenceFile>

   <!-- Loop over all Monte Carlo showers on file -->
   <loop numTimes="unbounded">

   <!-- Read in a Monte Carlo shower -->
   <module> EventFileReader </module>

      <!-- use each shower 10 times -->
      <loop numTimes="10" pushToStack="yes"> 

        <!-- Position the shower in random spot on simulated array -->
        <module> EventGenerator           </module>

        <!-- Simulate the surface detector response -->
        <module> ShowerRegenerator         </module>
        <module> G4TankSimulator           </module>
        <module> PhototubeSimulator        </module>
        <module> SdElectronicsSimulator    </module>
        <module> TankTriggerSimulator      </module>

        <!-- Simulate the fluorescence detector response -->
        <module> ShowerLightSimulator      </module>
        <module> LightAtDiaphragmSimulator </module>
        <module> TelescopeSimulator        </module>
        <module> FdBackgroundSimulator     </module>
        <module> FdElectronicsSimulator    </module>
        <module> FdTriggerSimulator        </module>

        <!-- Simulate the trigger, build and export event -->
        <module> CentralTriggerSimulator   </module>
        <module> EventFileExporter         </module>
      
      </loop>
   </loop>
</sequenceFile>
\end{verbatim} \end{small}

\caption{Example of a hybrid detector simulation module sequence.}
\label{f:HSim}
\end{figure}

The essential elements for the surface and fluorescence detector simulation
are contained within the innermost loop of the module sequence, while the outer loop 
allows for processing of all Monte Carlo events in a file or collection of files.
The surface detector simulation up to the tank triggering step is done
first, then the fluorescence simulation up to the local eye triggering is 
performed. The central triggering, event building and exporting are performed last.
It does not matter whether surface detector or fluorescence detector simulation is performed first, 
though both must be completed before the 
central triggering and event building steps occur.

Modularization allows one to easily substitute alternative
algorithms to perform a particular step of the simulation sequence. 
For instance, the detailed tank response simulation can be replaced
with a simplified, fast simulation by simply replacing the 
relevant {\tt <module>} element.  Such modularization of algorithms 
also allows collaborators to propose different approaches to a particular aspect of the
simulation process, and to compare results running under identical conditions.

\subsection{Hybrid Event Reconstruction}

\begin{figure}[htb]
\centering
\includegraphics[width=5.8in, height=5.6in]{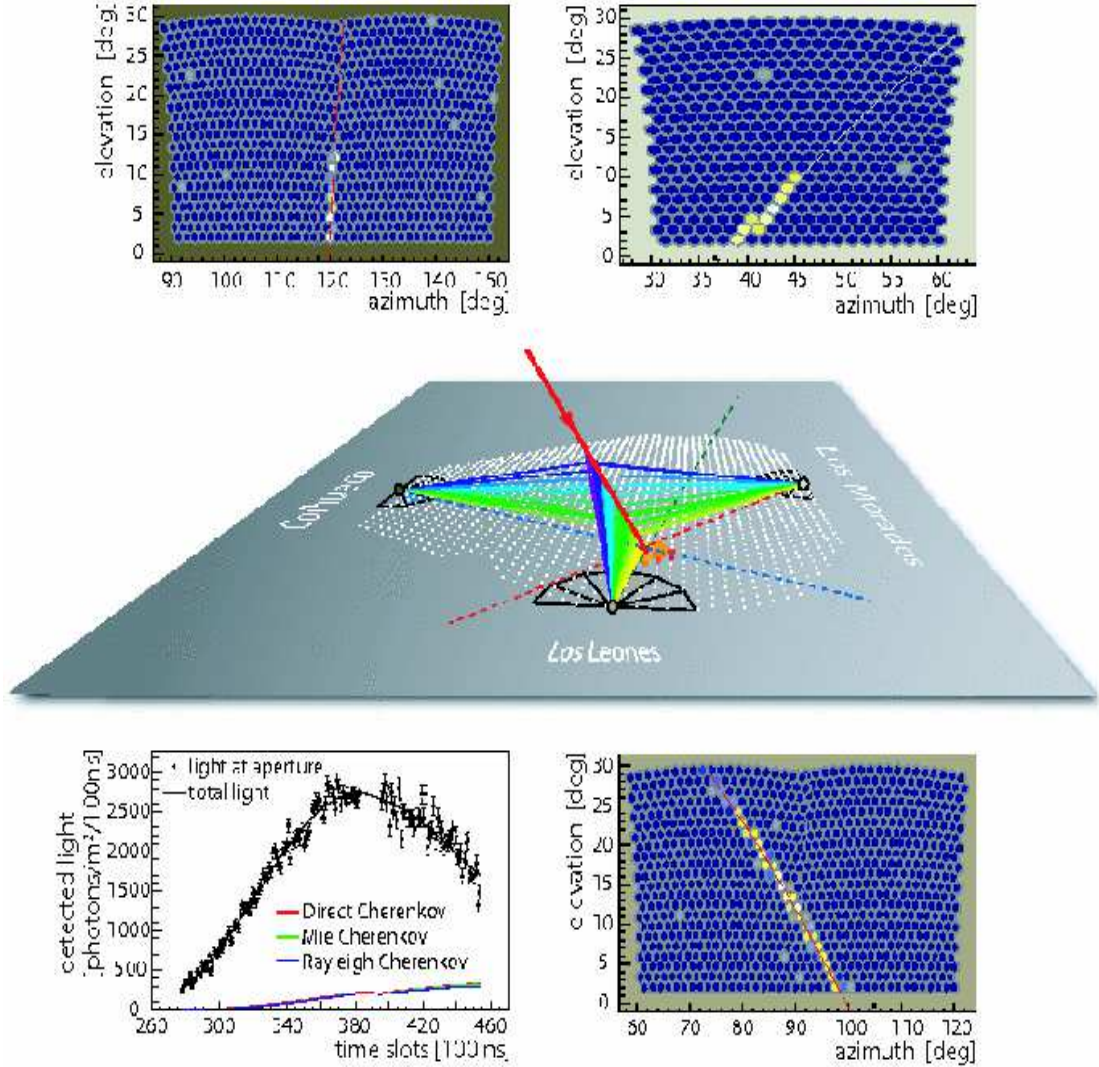}
\caption{Result of the hybrid reconstruction module sequence. 
The underlying event was created by simulations, 
and saved in the Auger Observatory raw data format. The figure shows the detector including 
the grid of 1600 surface detectors and the three (of four) fluorescence detectors which triggered
on the event. The 
colors (from blue to red) indicate the evolution of time. 
The three camera images show the image of the shower recorded on the telescope 
pixels, with the signal intensity on each pixel indicated by color.
Finally, the plot on the lower left shows the light profile arriving at the Los Leones telescope,
indicating contributions from different light sources.}
\label{f:reconstruction}
\end{figure}

The hybrid event reconstruction module sequence is indicated
in Fig.~\ref{f:HRec}.  This sequence begins with calibration 
of the fluorescence
and surface detectors, a procedure which transforms
real or simulated raw data into physical quantities. Afterwards 
a so-called pulse finding algorithm is used to further process the traces
recorded by the fluorescence telescopes.
Next, a series of geometrical reconstruction modules are employed. 
First the plane containing the shower axis and the eye which
detected it is determined.
A complete geometrical fit within this plane is performed, taking
into account both the timing of the shower image as it traverses the 
telescope pixels as well as the timing and impact point on the 
surface detectors.
A calculation of the light flux reaching the telescope aperture is then
carried out.
The last step is the profile reconstruction, which converts the 
fluorescence light profile recorded by the telescopes to a determination 
of the energy deposit at a given atmospheric depth along the shower axis.
The outcome of these steps is depicted in Fig.~\ref{f:reconstruction}.

\begin{figure}[htb]
\footnotesize
\centering
\begin{small} \begin{verbatim}
       ...
       
       <module> FdCalibrator           </module>
       <module> SdCalibrator           </module>    
       <module> FdPulseFinder          </module>

       <module> FdSDPFinder            </module>
       <module> HybridGeometryFinder   </module> 
       <module> FdApertureLight        </module>
       
       <module> FdProfileReconstructor </module>
       
       ...
\end{verbatim} \end{small}
\caption{Hybrid detector reconstruction module sequence.  This sequence fragment
can be appended to the one shown in Fig.~\ref{f:HSim} to reconstruct a simulated 
shower.}
\label{f:HRec}
\end{figure}

\section{Ongoing developments}

While the framework described in this note is actively used for analysis,
there are a few substantial enhancements in preparation.

First, we are developing an interactive visualization package which is
fully integrated into the framework and which will
provide not only graphical display of reconstructed event properties and
Monte Carlo truth, but also interactive control over configuration and 
reconstruction procedures.  This package will complement 
existing visualization tools which we use to browse 
processed events.  

Second, Python~\cite{python} bindings for the framework are in preparation.  
Once complete, all of the framework public interfaces will be 
exposed via Python, allowing 
users to prepare rapid prototypes of analysis and visualization tasks.
Python-based module sequencing will also be supported,
allowing more intricate run control than is currently afforded
through our XML-base sequencing system
for cases when this may be desired.  Although these bindings
will provide convenience for testing ideas and developing
algorithms, the existing module sequencing system and XML-based
run control will continue to be used for production runs, particularly 
as the logging features of this machinery is desirable for batch processing.

Third, the user module system described in section~\ref{sec:config} is 
being upgraded to support dynamical loading of modules.  This
will allow for easier use of modules with the interactive visualization
system mentioned above, and support easier module distribution and shorter
development cycles.

Finally, the event persistency machinery discussed in section~\ref{sec:event}
is undergoing revision.  Though the approach we have implemented
has been successful in decoupling the
in-memory event from the representation on disk, the design does 
impose a maintenance burden since any modification of 
the structure must be implemented both in the transient and persistent events.  
For the future we envisage a system employing a meta-description of the 
event which will be used to automatically generate the transient
and persistent events as well as the Python bindings mentioned above.
 
\section{Conclusions}

We have implemented a software framework for
analysis of data gathered by the Pierre Auger Observatory.
This software provides machinery to facilitate collaborative
development of algorithms to address various analysis tasks 
as well as tools to assist in the configuration and bookkeeping
needed for production runs of simulated and real data.
The framework is sufficiently configurable to adapt to a diverse
set of applications, while the user side remains simple enough
for C++ non-experts to learn in a reasonable time.  The modular
design allows straightforward swapping of algorithms for quick comparisons
of different approaches to a problem.  The interfaces to detector
and event information free the users from having to deal individually
with multiple data formats and data sources.  This software, while still 
undergoing vigorous development and improvement, has been 
used in production of the first physics results from the observatory.

\section*{Acknowledgment}

The authors would like to thank the various 
funding agencies which made this work possible,
as well as the fearless 
early adopters of the offline framework.




\begin{thebibliography}{}


\bibitem{Abraham:2004dt}
  J. Abraham {\it et al.} [Pierre Auger Collaboration],
  ``Properties and performance of the prototype instrument for the Pierre Auger
  Observatory,''
  Nucl. Instrum. Meth. A {\bf 523}, 50 (2004).

\bibitem{icrc2005}
S. Argir\`o {\it et al.} [Pierre Auger Collaboration],
  ``The offline software framework of the Pierre Auger Observatory,''
  FERMILAB-CONF-05-311-E-TD
{\it Presented at 29th International Cosmic Ray Conference (ICRC
  2005), Pune, India, 3-11 Aug 2005.}

\bibitem{xml}
{\tt http://www.w3.org/XML/};
Elliotte Rusty Harold, W. Scott Means,
``XML in a Nutshell'',
O'Reilly, 2004,
ISBN: 0-596-00764-7.

\bibitem{URI}
{\tt http://tools.ietf.org/html/rfc3986/.}

\bibitem{xpath} 
{\tt http://www.w3.org/TR/xpath}.

\bibitem{autotools}
{\tt http://www.gnu.org/software/autoconf};\\
{\tt http://www.gnu.org/software/automake};\\
{\tt http://www.gnu.org/software/libtool};
Gary V. Vaughn, Ben Ellison, Tom Tromey, Ian Lance Taylorm
``GNU Autoconf, Automake, and Libtool''
Sams, 2000,
ISBN 1-57870-190-2.

\bibitem{schema}
{\tt http://www.w3.org/XML/Schema/};
Eric van der Vlist, 
``XML Schema'', 
O'Reilly, 2002,
ISBN 0-596-00252-1.

\bibitem{md5} 
{R. Rivest, ``RFC 1321: The MD5 message-digest algorithm'',
{\tt http://www.faqs.org/rfcs/}}.

\bibitem{ice3-xml}
{See for example 
S. Patton, 
``Concrete uses of XML in software development and data analysis'',
International Conference on Computing in High-Energy Physics and Nuclear Physics 
(CHEP 2003), La Jolla, California, USA 24-28 March 2003.\\
{\tt http://www-conf.slac.stanford.edu/chep03/}.
}

\bibitem{d0framework}
J. Kowalkowski, H. Greenlee, Q. Li, S. Protopopescu, G. Watts, V. White, J. Yu,
``D0 offline reconstruction and analysis control framework'',
International Conference on Computing in High-Energy Physics and Nuclear Physics 
(CHEP 2000), Padova, Italy, 7-11 Feb 2000. \\
{\tt http://chep2000.pd.infn.it/}.

\bibitem{cms}
C.D. Jones, M. Paterno, J. Kowalkowski, L. Sexton-Kennedy, W. Tanenbaum,
``The new CMS event data model and framework'', 
International Conference on Computing in High-Energy Physics and Nuclear Physics
(CHEP2006), Mumbai, India, 13-17 Feb 2006.;\\
C.D. Jones,
``Access to non-event data for CMS'',
{\em ibid};\\
See also \texttt{http://cmsdoc.cern.ch/cms/cpt/Software/html/General/}.

\bibitem{icetray}
T. DeYoung,
``Icetray: A software framework for IceCube'',
International Conference on Computing in High-Energy Physics and Nuclear Physics
(CHEP2004), Interlaken, Switzerland, 27 September - 1 October 2004. \\
{\tt http://www.chep2004.org}.

\bibitem{aires}
{S. Sciutto, AIRES User's Manual and Reference Guide, \\
{\tt http://www.fisica.unlp.edu.ar/auger/aires}};
S.J. Sciutto,
  ``AIRES: A system for air shower simulations (version 2.2.0),''
  arXiv:astro-ph/9911331.

\bibitem{corsika} 
{D. Heck, J. Knapp, J.N. Capdevielle, G. Schatz, T. Thuow, Report FZKA 6019 (1998).}

\bibitem{conex}
T. Bergmann {\em et al.}, 
``One-dimensional hybrid approach to extensive air shower simulation'', 
Astropart. Phys. {\bf 26}, 420 (2007).

\bibitem{seneca}
H.J. Drescher, G. Farrar, M. Bleicher, M. Reiter, S. Soff, H. Stoecker,
``A fast hybrid approach to air shower simulations and applications'',
Phys.Rev. D{\bf 67}, 116001 (2003).

\bibitem{root}
{\tt http://root.cern.ch/.}

\bibitem{atlas-edm}
E. Moyse, F. Akesson,
``Event data model in ATLAS'',
International Conference on Computing in High-Energy Physics and Nuclear Physics
(CHEP2006), Mumbai, India, 13-17 Feb 2006. \\
{\tt http://www.tifr.res.in/~chep06/}.

\bibitem{mysql}
{\tt http://dev.mysql.com};
See for example Jon Stephens, Chad Russell,
``Beginning MySQL Database Design and Optimization:
From Novice to Professional''
Apress, 2004,
ISBN 1-59059-332-4.

\bibitem{xerces}
{\tt http://xml.apache.org/.}

\bibitem{stl}
{see for example N.\ Josuttis, 
``The C++ Standard Library'', Addison-Wesley, 1999, ISBN 0-201-37926-0.}

\bibitem{dtd}
{\tt http://www.w3.org/TR/REC-xml/}.

\bibitem{cppunit}
{\tt http://cppunit.sourceforge.net/doc/1.8.0/.}

\bibitem{buildbot}
{\tt http://buildbot.sourceforge.net/.}

\bibitem{boost}
{\tt http://www.boost.org/};
See for example Bj\"orn Karlsson, 
``Beyond the C++ Standard Library: An Introduction to Boost'',
Addison-Wesley, 2005,
ISBN 0-3211-3354-4.

\bibitem{clhep}
\texttt{http://proj-clhep.web.cern.ch/proj-clhep/.}

\bibitem{g4}
\texttt{http://geant4.cern.ch/};
S. Agostinelli \textit{et al.}, 
Nucl. Instrum. Meth. A \textbf{506}, 250 (2003);
J. Allison \textit{et al.},
IEEE Transactions on Nuclear Science, \textbf{53}, 270 (2006).

\bibitem{python}
{\tt http://www.python.org/};
Mark Lutz,
``Programming Python'',
O'Reilly, 2001, 
ISBN: 0-596-00085-5.


\end{thebibliography}
%




\end{document}